\documentstyle[aps,epsf,floats,preprint]{revtex}
\tighten
\begin{document}
\title{Finite dimensions and the covariant entropy bound}
\author{H. Casini}
\address{The Abdus Salam International Centre for Theoretical Physics, \\
I-34100 Trieste, Italy \\
e-mail: casinih@ictp.trieste.it}
\maketitle

\begin{abstract}
We explore the consequences of assuming that the bounded space-time subsets
contain a finite number of degrees of freedom. A physically natural
hypothesis is that this number is additive for spatially separated subsets.
We show that this assumption conflicts with the Lorentz symmetry of
Minkowski space since it implies that a conserved current
determines the number of degrees of freedom. However, the entanglement
across boundaries can lead to a subadditive property for the degrees of
freedom of spatially separated sets. We show that this condition and the
Poincare symmetry lead to the Bousso covariant entropy bound for Minkowski
 space.
\end{abstract}

\section{Introduction}

The quantum theory of fields suggests that an infinite number of degrees of
freedom live in the bounded space-time regions. To obtain a finite
dimensional Hilbert space one should consider energies bounded above by some
value in addition to a finite volume~\footnote{
A consistent definition of a Hilbert space of states assigned to some finite
region is at odds with the Reeh-Slieder theorem which implies that the field
operators in the space-time region generate the whole Hilbert space when
acting on the vacuum \cite{rs}. Even considering a bounded energy the notion
of a dimension for Hilbert space projectors assigned to space-time sets is
absent in the axiomatic versions of quantum field theory \cite{haag}.}.

However, there are reasons to suspect that the description in terms of a
quantum field theory (QFT) should break down below some distance scale. If
there is a fundamental cutoff the number of degrees of freedom could turn
out to be finite.

Finite Hilbert spaces are also a consequence of the picture coming from the
physics of black holes. These have an associated entropy equal to $A/(4G)$,
where $A$ is the horizon area and $G$ the Newton constant. If this number
can be included in the second law of thermodynamics \cite{bek}, the black
hole entropy would represent the maximum entropy of any system capable of
collapsing to form the black hole (assuming this later is stable 
\cite{wald}). Thought experiments of this kind have lead to the idea of the 
holographic
entropy bound \cite{st}. This means that the entropy of a system enclosed in
a given approximately spherical surface of area $A$ is less than $A/(4G)$.
An appropriate generalized version of this bound to arbitrary space-like
surfaces and general space-times was proposed by Bousso in 
\cite{bousso1,bousso}, where
it was called the covariant entropy bound. This is as follows. Given a
spatial codimension two surface $\Omega $ it is possible to construct four
congruences of null geodesics orthogonal to $\Omega $, two
past and two future directed. Suppose that one of these null
congruences orthogonal to $ \Omega $ has non positive expansion $\theta $ at
$\Omega $. Then, call $H$ the subset of the hypersurface generated by the
congruence that has non positive expansion. The hypersurface $H$ is called a
light-sheet of $\Omega $. The covariant entropy bound states that the
entropy in $H$ is less than $ A(\Omega )/(4G)$.

The covariant bound should be regarded as tentative. However, there are no
known reasonable counterexamples. Indeed, the bound can be shown to be true
in the classical regime under certain conditions equivalent to a local
cutoff in energy, and when the metric satisfies the Einstein equations \cite
{fmw}. This includes a vast set of physical situations.

Thus the entropy bounds would lead to a finite number of degrees of freedom
proportional to the bounding area. This is in apparent conflict with the
intuitive picture coming from a cutoff in space-time. If the dataset on a
Cauchy surface is arbitrary, this later roughly suggests that an independent
degree of freedom should be assigned to sets of the order of the cutoff
scale on these surfaces, leading to a number of degrees of freedom that
increases with the volume rather than the bounding area.

In this work we explore the consequences of assuming that a finite number of
degrees of freedom $n(A)$ can be assigned to bounded sets $A$ in
space-time. The function $n(A)$ can not be defined in QFT. We define it
here through a set of very general and physically motivated conditions.

We often use the term number of degrees of freedom in the sense of some
logarithm of the Hilbert space dimension $N$, what is proportional to the
statistical entropy $S=\log N$. For a system of independent spins $\log
_{2}N $ coincides with the number of spins~\footnote{
It is not necessary to leave continuity at this point. The dimension $N$
(and $\log (N)$) need not be discrete if we accept more general structures
than Hilbert spaces. Continuous dimension functions can be given to some
 lattices of projectors of von Newman algebras. These would be type type 
II factors in
contrast with the type I factors corresponding to Hilbert spaces and the type
III factors that form the algebras of operators of bounded regions in
algebraic quantum field theory \cite{haag}.}. The naive interpretation of the
dimension $N(A)$ is that it represents the number of independent states that
can be localized inside some Cauchy surface for $A$.

\section{Order and Causality}

The function $n(A)$ is a real function of space-time subsets. We will assume
that this function is an intrinsic property of the space. Evidently if a set
$A$ includes a set $B$, then it also contains all degrees of freedom that
under some reasonable definition can be said to belong to $B$. Thus, an
imperative condition for the function $n$ is that it preserves the inclusion
order between sets, that is

\begin{equation}
 \text{I- Order} \hspace{4cm} B\subseteq A\,\,\, \Rightarrow
\,\,\, n(B)\leq n(A) \hspace{3cm}
\label{order} \end{equation} for any $A$ and $B$.

The causal structure imposes that the physics inside the whole causal
development $D(A)$ of a set $A$ must be described in terms of the same
degrees of freedom. The causal development of a set $A$ in space-time is the
set of points through which any inextendible time-like curve intersects $A$.
A typical $D(A)$ set has the shape of a diamond (see Fig.1). Thus, the data
on the set $A$ for wave like equations determines the variables in the whole
set $D(A)$. This would imply that $n(D(A))\leq n(A)$, but taking into
account that $A\subseteq D(A)$ and (\ref{order}) we have $n(D(A))=n(A)$.

From now on we will take a more conservative domain for the
function $n(A)$, and focus attention on a special class of sets. These are
the ones that have a Cauchy surface~\footnote{
The function $D$ is sometimes defined only for sets having
a Cauchy surface \cite{wald1}.}. We call a Cauchy surface for a set $A$
to an achronal set $C$ included in $A$ such that $A\subseteq D(C)$. Thus,
we will
not consider the sets like two time-like displaced diamonds $U_{1}$ and
$U_{2}$. This is because it is possible that the function $n$ can not be
defined at all for these type of sets since the Cauchy problem is not well
defined on them. Evidently the data on the Cauchy surfaces for $U_{1}$ and
$U_{2}$ is necessary in order to solve a wave equation, but they are not
independent.

The argument we have given for the validity of a causal law for $n$ is in
terms of degrees of freedom. The same reasoning can be made for the number of
states.  If a state can be localized inside $D(A)$, meaning in a Cauchy
surface for $ D(A)$, then its future or past will pass through $A$, and the
converse is also true. This assumes that states cannot be created or
destroyed along their evolution and orthogonal states remain orthogonal.
Thus, underlying our second assumption is not only causality but
unitarity. The total number of independent states that can be localized on
the Cauchy surfaces $C_{1}$ and $C_{2}$ for the diamond shaped set in Fig.1
is the same. Then the second postulate reads

\begin{equation}
 \text{II- Causality + Unitarity} \hspace{1.5cm} n(D(C))=n(C)\ , \hspace{2cm}
\label{dos}
\end{equation}
for an achronal set $C$. The sets of the form $D(C)$ are called causally 
complete.

\begin{figure}[t]
\centering
\leavevmode
\epsfysize=4cm
\bigskip
\epsfbox{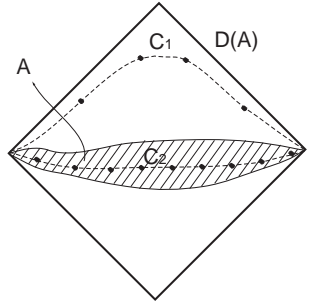}
\caption{The causal development $D(A)$ of the set $A$. The surface $C_{2}$ is
a Cauchy surface for $A$ while $C_{1}$ and $C_{2}$ are Cauchy surfaces for
$D(A)$. The points on $C_{1}$ and $C_{2}$ represent discrete degrees of
freedom on these surfaces (see the discussion at the end of Section III).}
\end{figure}

With the interpretation of $n$ as the
statistical entropy $\log N$ it may look strange to assign this number to
space-time subsets rather than to subsets of the phase space. However, in
theories for wave equations the causally complete sets uniquely determine
 subsets of the phase space through the initial data on their
Cauchy surfaces. This is not the case for sets without a Cauchy surface.

At least for globally hyperbolic space-times the bounded causally complete sets
 coincide with the bounded causally closed sets
\cite{ceg} (see Section V). Surprisingly, these form an orthomodular lattice, 
sharing this
property with the lattice of physical propositions in quantum mechanics
\cite{mio,ceg1}.

Thus, up to now we have only asked $n$ to be an order preserving function
 on the sets of the form $D(C)$ for some achronal set $C$. This is not very
restrictive.

\section{Additivity and a number of degrees of freedom proportional to the
volume}

Degrees of freedom located in space-like separated regions would be
independent since two operators based in them commute with each other. If we
think in a kind of lattice of spins on a Cauchy surface, the number of
independent spin degrees of freedom simply add for non intersecting subsets 
of the surface.
States for a union of spatially separated regions can be formed by tensorial
product, the dimension $N$ then increasing as the product of dimensions.
Based on this heuristic idea we can propose the following additivity
postulate
\begin{equation}
\text{III(a)- Additivity}\hspace{1.4cm} A\,\,\,\text{{\sl spacelike to}
}B\,\,\, \Rightarrow \,\,\, n(A\cup B)=n(A)+n(B)\,. \hspace{1.8cm} \label{tres}
\end{equation}

Equation (\ref{tres}) makes the function $n$ additive over subsets of a
Cauchy surface $C$ of a given causal diamond $U$. Thus, $n$ can be written as
the integral of a volume form $\omega _{C,x}$ on the surface $C$. We will
focus on strictly space-like Cauchy surfaces. The form $\omega _{C,x}$ at a
point $ x$ on $C$ depends on the Cauchy surface $C$ only locally, that is, it
is independent of how the Cauchy surface is extended out a neighborhood of
$x$ on $C$. Thus, a better nomenclature for the forms giving place to the
function $n$ is $\omega _{C,x}=z(x,\eta ^{\mu })\ \sqrt{g}\ \eta ^{\mu }\
\epsilon _{\mu \nu \alpha \beta }$, where $\eta ^{\mu }$, $\eta ^{\mu }\eta
_{\mu }=1$, is the normalized future directed time-like vector orthogonal to
$C$ at $x$. To be explicit, we can represent $n$ as
\begin{equation}
n\left( D(C)\right) =\int_{C}\ z(x,\eta ^{\mu }(x))\ \sqrt{g}\ \eta ^{\mu
}(x)\ \epsilon _{\mu \nu \alpha \beta }\,,\ \,  \label{equ}
\end{equation}
where $C$ is any space-like Cauchy surface for $D(C)$.

The combination of causality and additivity imposes severe restrictions to
the function $z(x,\eta ^{\mu }(x))$. To see this let us think in a diamond 
set $U$
of a differential size, so at zero order we can use the flat metric, and
assume the function $z$ is continue. Let us draw a Cauchy surface for
$U$ using pieces $C_{i}$ of
spatial planes passing through the diamond base faces, and let $\eta_{i}$ be
the corresponding normal vector to $C_{i}$. The number $ n(U)$ must not
depend on the chosen planes. Therefore, at zero order it is
\begin{equation}
\sum_{i} z(x,\eta _{i})\ v(C_{i}) = constant\,, \end{equation} where
$v(C_{i})$ is the volume of the piece of spatial plane $C_{i}$ we used for
constructing the Cauchy surface. Thus, this equation constrains the function
$z(x,\eta )$ for fixed $x$ and different $\eta $. The result is that the
function $z(x,\eta )$ must be of the form $j_{\mu }(x)\ \eta ^{\mu }$ for a
vector field $j_{\mu }$.  Then the integral in equation (\ref{equ}) is the
flux of a current over the Cauchy surface. As the number $ n(D(C))$ is
independent of $C$ the current $j$ must be conserved.

Resuming, we have that
\begin{equation}
n\left( D(C)\right) =\int_{C}j_{\mu }^{*}  \label{for}
\end{equation}
for a conserved $j_{\mu }$~\footnote{
Equation (\ref{tres}) makes the function $n$ a state on the
lattice of causally closed sets \cite{mio,ceg1}.
A full demonstration that a conserved current defines a state on the lattice
of causally closed sets though the formula (\ref{for}) can be consulted in
\cite{ceg2}.}.

Thus, it follows that the number of degrees of freedom comes from a
conserved current. This can not be obtained in a general space-time from the
metric alone. In particular it breaks the Lorentz symmetry of Minkowski
space.

Assuming finite dimensions, we have that either additivity or unitarity is
wrong, or Lorentz symmetry is broken. This seems to correlate with some
possibilities to regularize quantum field theories. For finite
regularization parameter the Pauli-Villars and the
higher derivatives regularizations give non unitary theories, and a
theory on the lattice is not Lorentz covariant.

The reason why additivity fails can be seen more easily looking at Fig.1.
Suppose we have a covariant way of assigning degrees of freedom to the Cauchy
 surfaces $C_{1}$ and $C_{2}$, for example taking the independent spin degrees
of freedom to be separated by some spatial distance. Then moving
$C_{1}$ to approach
the null boundary of the causal diamond its volume goes to zero reducing the
number of independent spins. As $C_{1}$ and $C_{2}$ must have the same number
 of degrees of freedom since they describe the same physics, most of the spins
 on $C_{2}$ must not be independent, and the prescription of separating the
independent spins by a fixed distance is wrong. However, the degrees of
freedom on the spatial corner may well turn out to be independent when
 restricting attention to a particular diamond (a related version of these
ideas is in Ref.\cite{mio2}).

\section{Subadditivity and the Bousso covariant entropy bound}

Therefore if we want to keep the postulates of Section II, and not to break
Lorentz invariance, we have to give up additivity.

Given two pieces $C_{1}$ and $C_{2}$ of a Cauchy surface $C$ such that $
C_{1}\cup C_{2}$ covers $C$, it is possible that some degrees of freedom are
shared between the $C_{1}$ and $C_{2}$, and then the number $n(C)$ will be
less than the sum $n(C_{1})+n(C_{2})$, while the opposite would be more
difficult to justify if the knowledge of the physics in $C_{1}$ and $C_{2}$
must determine the physics in $C$. Similarly, the entanglement of states for
adjacent regions in a Cauchy surface can lead to double counting the same
states \cite{tip}. The states on the boundary is counted by one or by
both, the states restricted to $C_{1}$ and the ones restricted to $C_{2}$.
Otherwise some states would escape from $C$. If $N(C_{1}).N(C_{2})\geq N(C)$
then $n(C_{1})+n(C_{2})$ $\geq n(C)$ would hold for the entropies. What
would be surprising is that such entanglement entropy could turn out to be
relevant.

Let us give place to the entanglement and assume that a subadditive law is
valid. If two sets $U_{1}$ and $U_{2}$ cover a Cauchy surface
for $U$ then
\begin{equation}
\text{III(b)- Subadditivity} \hspace{1.4cm}U\subseteq D(U_{1}\cup U_{2})
\,\,\, \Rightarrow \,\,\, n(U) \leq
n(U_{1})+n(U_{2})\ .\hspace{1.4cm}  \label{cuatro} \end{equation}

All requirements for the function $n$ up to now are conformal invariant. In
fact, they only depend on the causal structure rather than the conformal
one \cite{cau}. For example postulate (\ref{dos}) makes $n$ trivial for
some spaces without achronal sets.

Let us see more closely the implications of I, II and
 III(b) for the case of Minkowski space. The Poincare symmetry is not
shared by conformal transformations of the flat metric. Thus, imposing it
will select the Minkowski metric. This symmetry has very strong implications
for our function $n$. To see this let us take a small reference diamond $U$
with a cubic base on the surface $t=0$ in a given reference frame and with
faces of area $\alpha $. With a boost it is possible to stretch the diamond
along a null direction as in Fig.2. This stretched diamond must then have
the same number of degrees of freedom as the original one. The boosts do not
change the size of the transversal dimensions, and the area of the diamond
base spatial faces is of course unchanged. The null line that goes from the
diamond tip to the central point in the cubic face is orthogonal to that
face.

\begin{figure}[tb]
\centering
\leavevmode
\epsfysize=4cm
\bigskip
\epsfbox{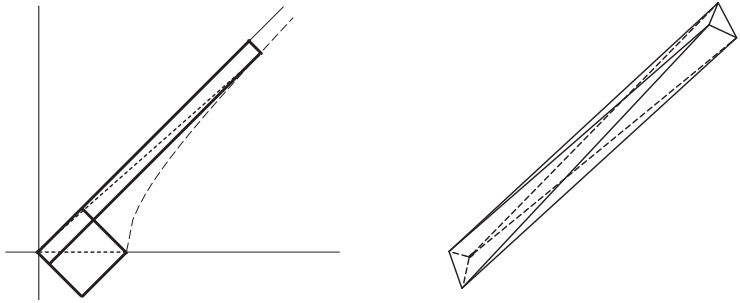}
\caption{Left: boosting a reference diamond. The short dashed line represents
the Cauchy surface formed by the diamond base. Right: a view of a
threedimensional stretched reference diamond.} \end{figure}

Consider now a space-like surface $\Omega $ of codimension $2$ and such that
its orthogonal null congruence has negative expansion at $\Omega $. With
many copies of boosted reference diamonds, using now rotations and
translations to align their spatial faces along the spatial surface $\Omega $,
 we can cover the whole light sheet of $\Omega $ (see Fig.3(a)). The
number of reference diamonds needed is the area of $\Omega $ in units of $
\alpha $. Each stretched diamond has the same number of degrees of freedom,
$n(U)$.  Then, using the subadditive property we have 
\begin{equation}
n(H)\leq \frac{n(U)}{\alpha }\ \ a(\Omega ),  \label{cota} 
\end{equation}
where $H$ is the light sheet of $\Omega $, and $a(\Omega )$ is the area of
the $(d-2)$ dimensional spatial surface $\Omega $. Note that the fraction on
the right hand side is independent of $\Omega $ as long as this surface is
big enough.  As shown in Fig.3(b), the same construction does not work for
expanding light sheets because the stretched diamonds separate each other
along the null congruence, and their transversal size can not increase since
boosts leave the transversal dimensions unchanged.

\begin{figure}[tb]
\centering
\leavevmode
\epsfysize=4cm
\bigskip
\epsfbox{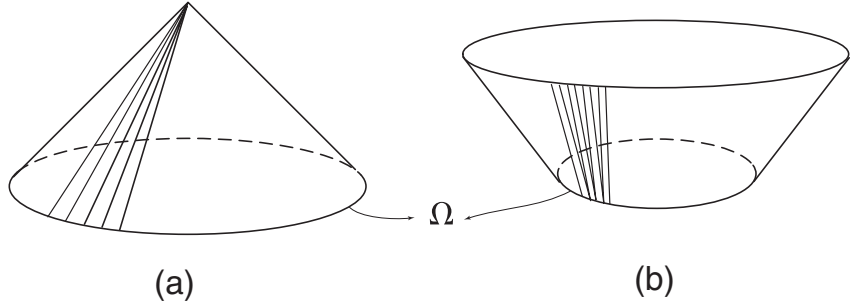}
\caption{(a) Covering a light-sheet with boosted reference diamonds. The
picture only shows their triangular null faces (in $d=3$). (b) It is not
possible to cover an expanding surface with stretched reference diamonds
since basing them on the surface $\Omega$ and aligning them along the null
lines they separate each other while their transversal size is constant.}
\end{figure}

Therefore, we recover all the elements of the Bousso entropy 
bound from
these simple assumptions. The constant relating area to entropy is not
determined at this point.

de Sitter space is conformally flat and it is also maximally symmetric with
symmetry group $O(4,1)$. Then we can use this symmetry group to emulate what
we have done for Minkowski space. To keep explicit the symmetry we can 
represent de Sitter space as the
hypersurface $x_{0}^{2}-(x_{1}^{2}+$ $x_{2}^{2}+x_{3}^{2}+x_{4}^{2})=-
\Lambda ^{2}$ in 4+1 dimensional Minkowski space. The Lorentz group 
$O(4,1)$ manifestly lives this surface
invariant. Let us consider the spatial surface $x_{0}=l$, and the sphere $
\Omega $ given by $(x_{1}^{2}+$ $x_{2}^{2}+x_{3}^{2})=r^{2}$, or
equivalently $x_{4}^{2}=l^{2}+\Lambda ^{2}-r^{2}$, on it. Among all the null
congruences orthogonal to the 2-sphere $\Omega $ in $5$ dimensions there are
 $4$
 included in de Sitter space.
These are also the $4$ null congruences orthogonal to $\Omega $ in de
Sitter space. Then, we can use the Lorentz boosts in $5$ dimensions as we
did before to stretch a small $4$ dimensional diamond covering a piece of $
\Omega $ along the null congruence orthogonal to it. These boosts must be in
a direction orthogonal to $\Omega $ but parallel to the hyperboloid at the
point where the small reference diamond is seated. Using rotations we can
transport diamonds along the spatial surface $x_{0}=l$. To transport the
reference diamond along the $x_{0}$ direction we have to use the boosts in
direction orthogonal to the hyperboloid at the point where the diamond is
located. The areas are unchanged by symmetry transformations  and we obtain 
again the bound
(\ref{cota}). As happens in the case of Minkowski space, this procedure can
work to obtain a bound as long as the surface is contracting. For positive
$l$ there is one future directed and one past directed light sheet of $\Omega
$ for $r<\Lambda $, while for $r>\Lambda $ there are two past directed
light-sheets.

For spaces with less number of symmetries it would not be possible to obtain
the covariant bound using a single reference diamond as we have done here. In
the general case, given a causal structure we have to study the possible
solutions to the set of constraints I, II and III(b). 
We remark that this system of inequalities is linear, thus
linear combinations with positive coefficients form also solutions. We will
not develop further in this sense here. Instead we go back to the case of
Minkowski space.

\section{The scaling Newton constant}

The equation (\ref{cota}) compared with the covariant entropy bound suggests
the interpretation of $\frac{4n(U)}{\alpha }$ as $G^{-1}$. However, the
value of $\frac{n(U)}{\alpha }$ must be used only for the light-sheets $H$
with $a(\Omega )$ greater than $\alpha $. The smallest value of $\frac{n(U)}{
\alpha }$ for the reference diamonds $U$ adequated to a given $H$ is what 
should be used to have
the strongest bound.

The construction of the preceding Section can be generalized to investigate
the scaling properties of $n(U)$ in a way similar to the renormalization
group, but where we have inequalities instead of equations. The order
and subadditivity play opposite rolls in the inequalities, and the Poincare
symmetry is the tool to use for comparing different sets. The equation (\ref
{cota}) is obtained by measuring a set using another one and is a special
case of these inequalities. Many other configurations could be used to
provide different inequalities of this kind.

For example consider diamonds with spatial rectangular base at $t=0$ in a
given reference frame. Let us call them $U_{(l_{1},..,l_{d-1})}$ where $
(l_{1},..,l_{d-1})$ are the side lengths. Here $d$ is the space-time
dimension. The diamond $U_{(l_{1},..,l_{d-1})}$ can be measured by two
copies of the diamonds $U_{(l_{1},..,\varepsilon )}^{1}$,.., $
U_{(l_{1},.,l_{k-1},\varepsilon ,.,l_{d-1})}^{k}$,.., $U_{(\varepsilon
,.,.,l_{d-1})}^{d-1}$, where $\varepsilon $ is smaller than all the other 
 sides. It is
enough to cover with the big faces of the $U_{(l_{1},.,l_{k-1},\varepsilon
,.,l_{d-1})}^{k}$ the corresponding faces of $U_{(l_{1},..,l_{d-1})}$ and
make the boosts in the small direction (the $k^{th}$ direction).
The subadditive and order properties give
\begin{eqnarray}
\max (n(U_{(l_{1},.,l_{k-1},\varepsilon ,.,l_{d-1})}^{k}))\leq
n(U_{(l_{1},..,l_{d-1})})\leq 2(\sum n(U_{(l_{1},.,l_{k-1},\varepsilon
,.,l_{d-1})}^{k})) \nonumber 
\\  \leq 2(d-1)\max (n(U_{(l_{1},.,l_{k-1},\varepsilon
,.,l_{d-1})}^{k}))  \label{capa}
\end{eqnarray}
There we see that $n(U_{(l_{1},..,l_{d-1})})$ is roughly given by the
function $n$ of its facial diamonds $U^{k}$.

Let us choose $l_{k}=l$ smaller than all the other sides,
which makes the face perpendicular to this side
 the biggest one.
Then we see from (\ref{capa}) that the function $n(U_{(l_{1},.,l_{k-1},
\varepsilon ,.,l_{d-1})}^{k})$ as $\varepsilon $ goes to zero is bounded
below by $n(U_{(l_{1},..,l_{k-1},l,..l_{d-1})})/(2(d-1))$. It also decreases
with $\varepsilon $ and thus must converge. We call to this limit for facial
sets
\begin{equation}
n_{d-2}(l_{1},..,l_{d-2})=\lim_{\varepsilon \rightarrow
0}n(U_{(l_{1},..,l_{d-2},\varepsilon )}).
\end{equation}
We have only rotations and translations as symmetries for the facial sets,
since we have
run out of boosts. The function $n_{d-2}$ is ordered by inclusion. The
subadditive law implies
subadditivity for faces, that is, the sum of the function $n_{d-2}$ of the 
faces that
cover a given face must be greater than the function $n_{d-2}$ for this face.

Covering with copies of a face $(r_{1},...,r_{d-2})$ another face $
(l_{1},..,l_{d-2})$ with multiple side lengths, $l_{k}=m_{k}r_{k}$,
$m_{k}\geq 1$ integer, we obtain \begin{eqnarray}
\frac{n_{d-2}(l_{1},..,l_{d-2})}{l_{1}...l_{d-2}} &\leq &\frac{
n_{d-2}(r_{1},...,r_{d-2})}{r_{1...}r_{d-2}},  \label{esta} \\
n_{d-2}(r_{1},...,r_{d-2}) &\leq &n_{d-2}(l_{1},..,l_{d-2}).\label{otra}
\end{eqnarray}
Thus, the $(d-2)$ dimensional function $\tilde{G}
_{d-2}^{-1}=(4n_{d-2}/A_{d-2})$, where $A_{d-2}$ is the surface area, is a 
decreasing function in the sense of the
inclusion order, while  $(A_{d-2}.\tilde{G}_{d-2}^{-1})$ is an increasing
function. As a result the behavior of $\tilde{G}_{d-2}^{-1}$ is bounded
between a constant and a constant times $A_{d-2}^{-1}$.

The area law for entropy appears here as a possibility unrelated a priori
to gravity.
The interpretation of $\tilde{G}_{d-2}$ as proportional to the
Newton constant is supported by the work \cite{jac}. There the entropy
was taken proportional to the area and
 the stress tensor was interpreted in terms of a heat flux across null
surfaces. Using
the second law of thermodynamics as seen from accelerated observers to relate 
the
changes of area with the heat flux it was shown that the Einstein equations
for the metric can be obtained. It also follows the proportionality
factor $1/4$ between $G^{-1}.area$ and entropy. We think that a rewriting of
 that work in terms of the function $n$ and where the stress tensor
provides additional information to the inequalities could clarify the roll of
gravity in the present context.

We note two things. The first is that according to (\ref{capa}) the ratio $
\frac{n(U)}{A(\Omega _{U})}$, where $\Omega _{U}$ is the surface forming the
spatial corner of $U$, is of the order of $\tilde{G}_{d-2}^{-1}$ for an area
of the same size. The second is that $\tilde{G}_{d-2}^{-1}$ can depend on the
size and form of the surface. Gravity could be stronger at larger size
according to (\ref{esta}).

We will leave a more complete analysis of the possible solutions of the
system of inequalities for a future work. But we note that there are two
solutions for $n$ that are extremal in the sense explained before. These are
$n(U)=c_{1}$ for a constant $c_{1}$ and $n(U)=c_{2}A(\Omega _{U})$. In this
later case we have to consider only the sets $U$ whose null border has
positive expansion at $\Omega _{U}$ (see the next Section).
A linear combination of solutions with positive coefficients is also a 
solution, so
$n(U)=c_{1}+c_{2}A(\Omega _{U})$ is a solution. In this case we have
\begin{equation} \tilde{G}_{d-2}=\frac{1}{4(c_{2}+\frac{c_{1}}{A_{d-2}})}.
\end{equation} The case of $c_{2}=0$ corresponds to a number of degrees of 
freedom that 
saturates and does not increase any more with the size. All sets share the same
degrees of freedom and the entanglement is total. The case $c_{1}=0$ is when
the Newton constant is really constant and the area law for the entropy is
valid.

In general both types of solutions can contribute, and then the picture that
emerges is that as we go to greater sizes the term corresponding to a number
of degrees of freedom proportional to the area dominates and 
$\tilde{G}_{d-2}$ is constant. This is a standard gravity with constant $G$. 
At lower sizes
the constant number of degrees of freedom could dominate and the Newton 
constant would 
vanish. The Plank scale would increase as we go to smaller scales making
impossible to surpass it and leading to a regularization for all other fields
 \cite{haba}. 
This limit is conformally invariant since the
number of degrees of freedom is constant. 

In a general space-time $n$ has to be assigned to diamonds $U$, and being
subadditive it is less than the sum of a number proportional to the spatial
corner area of small stretched diamonds that cover the null border of $U$.
Thus we see a reason why $n$ would be an additive function on the spatial
border of $U$ for big sets. This is because it is the most relevant term in
the language of the renormalization group since all other terms have to be
subadditive. The area is the less suppressed of the additive terms on the
spatial corner when the curvature is small. Thus, it is possible to imagine
that given a causal structure the solution of the inequalities would give an
area law, from which to extract the metric. This later would result composed 
by two 
parts, a net of causal diamonds corresponding to some lattice of projectors, 
and the function $n$, corresponding to the logarithm of a dimension function 
for the projectors. 
The metric would have no sense
for smaller diamonds with entropies that are non additive over the spatial 
corner.

However, as the basic element is the causal structure and not the metric, it
seems that a different interpretation for the conformal case $n=constant$ is
possible. This implies that the symmetry under scaling of the causal 
structure of Minkowski space is realized. Thus, it is possible that a
different metric could realize this symmetry in the sizes where $n$ is 
constant. A sector of de
Sitter space with the metric
\begin{equation}
ds^{2}=\frac{\Lambda ^{2}}{t^{2}}(dt^{2}-dx^{2}-dy^{2}-dz^{2})
\end{equation}
is invariant under scaling $x_{\mu}^{\prime}=\lambda x_{\mu}$. 
Time translation symmetry is however broken.

\section{The domain of the number of degrees of freedom function}
According to the causal postulate II the domain of $n$ can be taken as the 
bounded causally complete sets. As we have mentioned these sets coincide with 
the bounded causally closed sets \cite{ceg,mio,ceg1}. These are the sets $S$ 
satisfying $S=S^{\prime \prime}$ where the spatial opposite of $S$, 
$S^{\prime}$, is the set of points that can not be connected by a time-like 
curve with any point in $S$. 

The non expanding condition for the null border of the causally closed sets 
$U$ would imply a new restriction to the domain of $n$, since causally closed
 sets where some piece of the null border is expanding at the spatial corner
 could not have a well defined $n$. 

Surprisingly, it seems that this new restriction can be implemented by 
imposing that $S$ is not only closed under the double spatial opposite but
 also under twice the time-like opposite operation. That is we would have to 
consider sets $S=S^{**}$, 
where $S^{*}$ is the set of all points in space-time that can be connected by 
a time-like curve with all the points in $S$. We will call to these sets the 
observable sets. The reason for this name is that if an observer can see the 
whole set $S$ then it must see also all $S^{**}$. The set $S^{**}$ is the 
intersection of the sets that can be seen from the observers that can see $S$.
 In this sense an observer that see two points spatially separated in 
Minkowski must also see the geodesic path joining them, so the points can not 
be separated from the joining path. Observers here are 
taken in a time symmetric sense. The restriction to the observable sets, 
$S=S^{**}$, is a generalization of the ideas in \cite{bousso3}. There it was 
argued that the sets formed by the intersection of the past of a point in an 
observer 
world-line and the
 future of another point in the same world-line (causal diamonds) are the 
actual sets accessible to observation. We see that our definition for the 
observable sets allow also to information exchange between observers.  
       
If a set is observable or not is calculable with the only knowledge of the 
causal structure. However, the non expanding condition is non conformally
 invariant. Then these conditions are not equivalent in any space-time. It
 is possible that the Einstein Equations with some energy condition would 
lead to the equivalence of being observable and having non expanding null 
border at the spatial corner. In fact for a spherical surface $\Omega$ 
greater than the horizon in a Robertson Walker model the corresponding diamond 
set $U$ with spatial corner $\Omega$ does not have positive expansion at 
$\Omega$. However, the presence of the singularity makes $U^{**}$ a set 
without spatial corner. 

\section{Acknowledgments}
I have benefited from correspondence with R. Bousso. I thank
specially M. Huerta for encouragement and discussions. This work was partially
supported by the Abdus Salam International Centre for Theoretical Physics.

\end{document}